\documentclass[prb,showpacs,nofootinbib,twocolumn,amsmath]{revtex4}
\usepackage{hyperref}
\usepackage{bm,amsfonts,mathtools}
\usepackage{graphicx, wasysym}
\usepackage{textcomp}
\usepackage{amssymb,xcolor,latexsym,epsfig}
\newcommand{\be}{\begin{equation}}
\newcommand{\ee}{\end{equation}}

\usepackage{epstopdf}
\usepackage{hyperref}
\usepackage[utf8]{inputenc}
\usepackage{xcolor}
\def\bfk{{\bf k}}

\def\bfb{{\bf b}}
\def\be{\begin{equation}}
\def\ee{\end{equation}}
\begin{document}
\vspace*{0.5cm}
\title{Ansatz about a zero momentum mode in QCD and the forward slope in pp elastic scattering\footnote{This article is written in memory of our dear friend and collaborator Rohini M. Godbole (1952-2024), an extraordinary physicist and a mentor of women scientists, with whom we shared a common passion for understanding QCD effects in  hadron and photon collisions.} }
\author{G. Pancheri} 
\affiliation{INFN, Laboratori Nazionali di Frascati, 00044 Frascati, Italy}
\affiliation{Centro di Ricerche Enrico Fermi, Roma, 00185, Italy }
\author{Y. N. Srivastava }
\affiliation{Centro di Ricerche Enrico Fermi, Roma, 00185, Italy }
\affiliation{Emeritus Professor, Northeastern University, Boston, USA}
\author{O. Shekhovtsova}
\affiliation{INFN Sezione di Perugia, 06123 Perugia, Italy}
\affiliation{National Science Centre, Kharkov Institute of Physics and Technology, 1
Akademicheskaya, Ukraine}
\begin{abstract}
We recall a resummation procedure in QED to extract the zero momentum mode in soft photon  emission and present an ansatz about a possible mechanism for the forward peak characterizing elastic proton proton scattering.
\end{abstract}
\maketitle
\section{Introduction}
The observation of the rise of the total proton-proton  cross-section at the {Intersecting} Storage Rings (ISR) accelerator \cite{Amaldi:1972uw} was one of the earliest signals of non scaling phenomena  in hadronic physics, as  surprises  arrived when  the increasing   proton c.m. energy passed the threshold between quark confinement and asymptotic freedom.  The rise had been anticipated by cosmic ray observations \cite{Yodh:1972}, and was  among other unexpected results, such as the excess in multi hadron production in electron-positron collisions, first observed  at ADONE, when it started its operation in 1969 with $\sqrt{s}\approx 1.6- 2 $ GeV \cite{Bartoli:1970df}.  The observation was soon confirmed at the Cambridge Electron Accelerator and later at SPEAR,  showing that a threshold had been passed  as it became quantitatively evident  a few years later,  in November 1974, with  the discovery of 
a new particle, later called the  $J/\Psi$-meson. It was  a bound state of a new quark, the {\it charm}, a very narrow resonance with $3.1$ GeV mass {\cite{Aubert:1974js,Augustin:1974xw,Bacci:1974za}}, 
{an   energy at which the strong coupling constant can be expected to be small enough to allow a perturbative behaviour \cite{DeRujula:1974rkb}. }

Similarly, in hadron interactions, the observation of the rise of the total $p-p$ cross-section can be expected  when  the c.m. energy for  parton-parton  collisions is  around 2-3 GeV,   which would correspond to $E_{cm}^{pp}\approx 20$ GeV, in a simple model where  each quark in the proton carries 1/6 of the energy \cite{Fagundes:2015vba}.
Indeed, a  rising behaviour, which one can  attribute to semi-hard collisions,  had been reported to appear  in cosmic rays experiments in 1972 \cite{Yodh:1972} and was confirmed when the ISR  \cite{Amaldi:1971kt,Amaldi:1972uw} started taking data for the total and elastic cross-sections. Further experimental studies of the elastic differential cross-section gave evidence for structure   in $ pp$ collisions. As to  the detailed mechanism for the rise of $\sigma_{total}$, we have long advocated  for the rise  being  a collective effect  of mini-jet production, accompanied by soft gluon resummation. Namely, the rise is due to the appearance of interacting quarks and gluons and it is modulated by the unavoidable soft gluon emission.The problem of such model to be quantitative  is the still not understood behaviour of the strong interaction coupling for  very small momentum, near zero, soft gluons. 

In previous publications \cite{Godbole:2004kx}, we have been able to describe  the energy dependence of the total $pp/{\bar p}$   cross-section \cite{Pancheri:2016yel} through currently available  LO PDF for parton parton collisions, and an infrared safe resummation procedure with a singular but integrable coupling constant $\alpha_s(k_t) $ such that
\begin{eqnarray}
\alpha_s^{IR} (k_t)=\frac{12 \pi}{11 N_c-2N_f} 
(\frac{k_t^2}{\Lambda^2}
)^{-p}  \ \ \ \ \ p< 1 \label{eq:alphasIR}\\
\alpha_s^{AF} (k_t)=\frac{12 \pi}{11 N_c-2N_f}\log [\frac{k_t^2}{\Lambda^2}] \label{eq:alphasAF}
\end{eqnarray}
  where Eq.~(\ref{eq:alphasIR}) ensures infrared integrability for the resummation procedure, and Eq.~(\ref{eq:alphasAF}) appropriately describes perturbative behaviour for the mini-jet collisions. In our  published phenomenology  \cite{Godbole:2004kx,Achilli:2011sw}
  we used an interpolating expression 
  \begin{equation}
  \alpha_s^{BN}(k_t)=\frac{12 \pi}{11 N_c-2N_f}
  \frac{p}{
  \log[
  1+p(
  \frac{k_t}{\Lambda}
  )^{2p}
  ]
  } \label{eq:aphaBN}
  \end{equation}

\begin{figure}
 \centering
 \includegraphics[width=8.5cm]{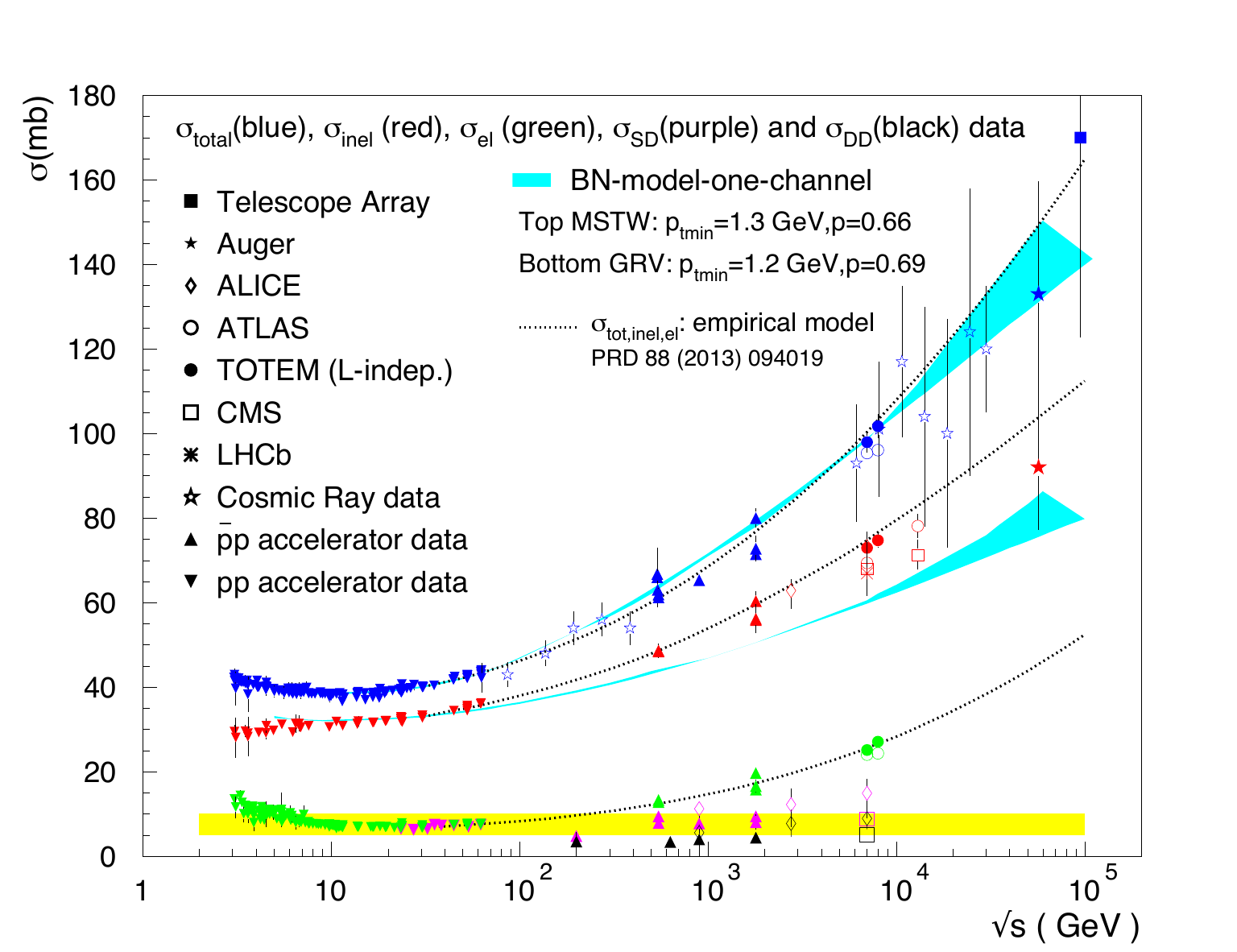}
 \caption{The figure \cite{Pancheri:2014rga} shows the total, elastic and inelastic cross-section in  the BN model (blue band)\cite{Godbole:2004kx}, 
   and an empirical model  (dots) \cite{Fagundes:2013aja}. The yellow band is the diffractive contribution from data. }
 \label{fig:BNmodeltotalxs}
 \end{figure}

We called BN model our proposal for the rising total  pp cross-section, from the Bloch and Nordsieck \cite{Bloch:1937pw}  theorem, which inspired Touschek's soft photon resummation procedure \cite{Etim:1967mxo}, which we have extended to treat  soft gluon emission during the semi-hard collision. Such a model was based on  eikonal resummation of QCD  mini-jets \cite{Durand:1988cr}, with an impact parameter distribution inspired by our previous work on soft gluon effects in hadronic collisions \cite{PancheriSrivastava:1976tm}. The BN model    reproduced total cross-section data up to LHC energies and available cosmic ray data within reported errors,  but unlike other models \cite{Jenkovszky:2023wev,Khoze:2018kna,Luna:2024cbq}
 could  not   reproduce the elastic  or quasi-elastic  process, and {underestimated the  total inelastic cross-section,  as shown in Figs~(\ref{fig:BNmodeltotalxs})  \cite{Pancheri:2014rga}. This difficulty  could be due to our model, based  so far on a single  channel component. Presently, we refrain from going  beyond  the one channel \cite{Ryskin:2009tj}, as this would  introduces   extra  parameters.}
 As for  the differential cross-section, the BN model did not show a  dip structure (only bumps and zeroes) \cite{Jenkovszky:2024avi} 
 nor an exponential behaviour for  $-q^2\approx 0$,   as observed in  the    forward peak \cite{Fagundes:2016rpx}. {Such behaviour can be   obtained in  Regge-type models 
 but did  not naturally  arise in our  BN model. }

\begin{figure}
 \includegraphics[width=8.5cm]{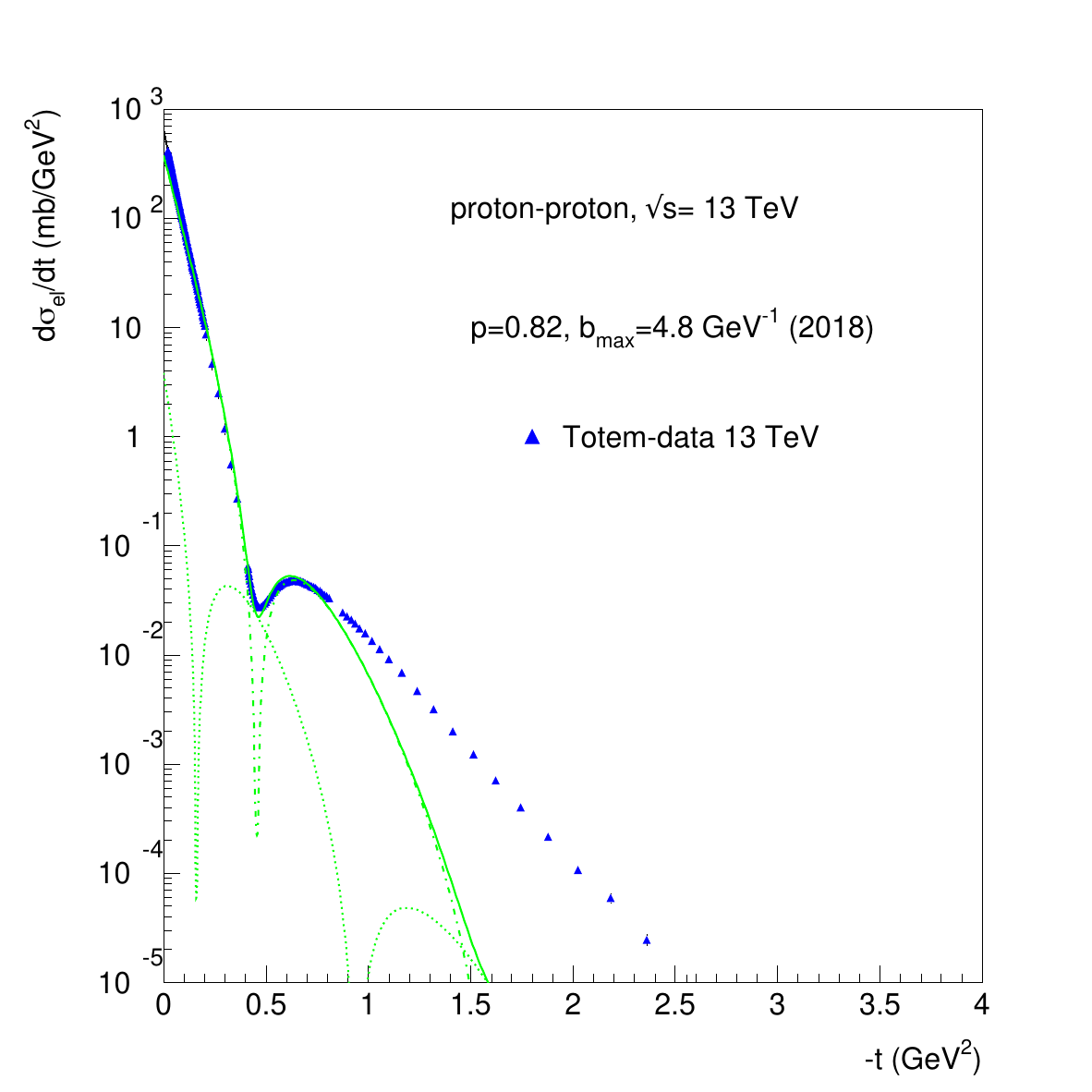}
 \caption{TOTEM 13 TeV data and the BN model with an   exponential cut-off, plot courtesy from  A. Grau.}
\label{fig:bn13TeV}
\end{figure}

 To test the forward region,   an empirical    $bmax$ can be introduced to multiply   the differential  cross-section from our model  at $\sqrt{s}=13$ TeV,   obtaining a  good description of data \cite{TOTEM:2017asr} up to the {\it bump}, as shown in Fig.~\ref{fig:bn13TeV}.

In order  to understand the origin of a cut-off  in  QCD in the context of our model,  we now turn   to address the question of zero momentum modes in gauge theories \cite{Palumbo:1983bf}.
After summarizing both the Touschek resummation method in QED and its extension to deal with zero momentum modes,   we will propose an ansatz for the origin of the forward peak, based on a revisitation of the Touschek's procedure.
  The ansatz we present has  a correspondance with   Color Condensate Models,  
  but differs in that   we derive its origin  from the zero mode in soft emission processes resummed through the Touschek method \cite{Etim:1967mxo,Palumbo:1983aa}.
   
    In Sect.~\ref{sect:zeroQED} we  recollect the results from previous  work  about the zero momentum mode in Abelian gauge theories. We then extend the discussion to the transverse momentum distribution in QCD  and show how hadronic processes could exhibit a forward peak in the elastic differential cross-section, arising from the zero momentum mode. We shall then have a discussion about the proposed expression for the interpolating $\alpha_s$, Sect.~\ref{sec:alphas}, leaving the relevant phenomenology to a further publication. The article will  close with some considerations about the strong coupling constant in its two asymptotic limits. 

 \section{Zero momentum mode in Abelian gauge theories}
 \label{sect:zeroQED}
 
Our argument  would follow  two papers  \cite{Palumbo:1983aa,Palumbo:1983bf}  about the connection between boundary conditions in gauge theories and its implications following the possible existence of zero momentum modes.   The question had been   posed whether unexpected physical effects can rise  
in the transition from a sum over discrete modes to continuum distributions  \cite{Palumbo:1983bf}. Using the Fourier expansion of the gauge fields, it was argued  that the gauge field should vanish at the boundaries of the QED quantization box, whereas 
 the question arises whether periodic boundary conditions in QCD might give  observable effects. The problem was rediscussed \cite{Palumbo:1983aa},
 using  Bruno
 Touschek resummation formalism  \cite{Etim:1967mxo},  EPT  for short,  based on a semi-classical  approach to calculate  the    probability distribution of soft photons, emitted in charged particles collisions. 


 \subsection{Touschek's formalism for QED} Touschek's  objective was to calculate the probability of unobserved  soft photon emission up to an experimental resolution $\Delta \omega\le E$, the scale of the process and obtain the correction factor to the measured electron-positron 
 cross-section. Touschek started with  the Bloch and Nordsieck result about  soft photon emission from a classical source \cite{Bloch:1937pw}. Bloch and Nordsieck  had shown that the distribution in the number of photons was given by a Poisson distribution, hence the probability of emission of  soft photons with different values for the momentum {\bf k} would be given by the product of their Poisson distributions,  namely
 \begin{equation}
 P( \{n_{ \bf k }\}) =\Pi_{\bf k}
 \frac{
{\bar n}_{\bf k}^{n_{\bf k}}
 }
 {n_{\bf k}!} \rm{exp}[-{\bar n}_{\bf k}]
 \label{eq:Poisson}
 \end{equation}
with $n_{\bf k}$    the number of  photons emitted with  momentum ${\bf k} $ around their average value   ${\bar n}_{\bf k}$.
We notice that Eq.~(\ref{eq:Poisson})
describes a discrete momentum spectrum  of the emitted photons, corresponding to quantization of the electromagnetic field in a finite box. In the following, we shall first assume that a smooth continuum limit exists. As we go through  Touschek's argument,  we shall also point out possible subtleties with the continuum limit.     

 The next four steps taken by Touschek are:
\begin{enumerate} 
\item sum over all  values of number of soft photons of momentum {\bf k}, namely  
\begin{equation}
 \sum_{n_{\bf k}}P(\{n_{\bf k}\})
\label{eq:sumpoisson}
\end{equation} 
\item the  probability of having a 4 -momentum loss  between $K_\mu$ and $K_\mu +d^4K$ 
 due to   all  possible number  of emitted photons $n_{\bf k}$ and all possible    single  photon momentum {\bf k}, is obtained by imposing 
 overall energy momentum conservation, through the function  $\delta^4(K_\mu-\sum_{\bf k} n_{\bf k}k_\mu)$ expressed in its $\mathcal{F}ourier$ trasform
 \begin{equation}
 \label{eq:delta}
 \delta^4(K_\mu-\sum_{\bf k} n_{\bf k}k_\mu)= \frac{1}{(2\pi)^4}\int d^4x e^{-i(K_\mu-\sum_{\bf k} n_{\bf k}k_\mu)\cdot x}
 \end{equation}
\item  one can then exchange the product of the infinite number of Poisson distributions, Eq.~(\ref{eq:Poisson}) with the sum in Eq.~(\ref{eq:sumpoisson}),
\item and take the continuum limit, {unless there are special boundary conditions, as discussed in the next section.}
\end{enumerate}
In the following we shall refer to this work as EPT paper. Explicitly, the above steps are implemented  as
\begin{eqnarray}
d^4P( K)=\sum_{n_{\bf k}}P(\{n_{\bf k}\}) d^4{ K}  \delta_4( K-\sum _{ \bf k}  k n_{\bf k})= \nonumber\\
\sum_{n_{ k}}  \Pi_{\bf k}
 \frac{[\bar{n}_{\bf k}]^{n_{\bf k}}}{n_{\bf k}!}e^{-\bar{n}_{\bf k}} d^4{ K}  \delta_4({ K}-\sum_{\bf k} { k} n_{\bf k}) \label{eq:d4P}
\end{eqnarray}
Touschek proceeds to steps 2 \& 3 by  using the integral representation of the delta-function to exchange    the sum with
the product obtaining
\begin{equation}
d^4P( K)=\frac{d^4  K} {(2\pi)^4}
\int d^4 x  e^{ -i{ K \cdot x } }exp\{   -\sum_{\bf k}{\bar n}_{\bf k}  [ 1-e^{  i  k\cdot x} ]   \}
\label{eq:prob4discrete}
\end{equation}
Going to the continuum, brings
\begin{equation}
d^4P( K)=\frac{d^4 K}{(2\pi)^4}\int d^4 x e^{-i  K \cdot x } exp\{-\int d^3{\bar n}_{\bf k}
[1-e^{i k \cdot x } ]
\} \label{eq:prob4continuum}
\end{equation} 
In this formulation, an important property of the integrand in Eqs.~(\ref{eq:prob4discrete}) and  (\ref{eq:prob4continuum}) is that by its definition $d^4 P(K)\ne 0$ only for $K_0\equiv \omega \ge 0$, since for each single photon $k_0\ge 0$ .

\subsection{The continuum limit and the closed form expression for the energy distribution}
If one takes  the continuum limit, integrating over the three momentum $\bf K$ leads to the  probability of finding an  energy loss in the interval $d\omega$ as
\begin{eqnarray}
\label{eq:dpomega}
dP(\omega)= \frac{d\omega}{2\pi}\int_{-\infty}^{+\infty} dt \ \rm{exp[i\omega \ t  - h(t)] }=\nonumber\\
\frac{d\omega}{2\pi}\int_{-\infty}^{+\infty} dt \ \rm{exp[i\omega \ t  - \beta(\alpha_{em}; p_i^\mu) \int_o^{\epsilon_s}\frac{dk}{k}(1-e^{-ik t}) ] }
\end{eqnarray}
where one has used the property of separation between the angular and the momentum integration over the photon momenta,  already exploited by Weinberg \cite{Weinberg:1965nx},
and known to previous authors as well.
The separation defines   $\beta(\alpha_{em};p_i^\mu)= \beta$ as a function of the incoming and outgoing particle momenta $(p_i^\mu)$, i.e.
\begin{equation}
\beta=\frac{\alpha}{(2\pi)^2} \int d^2 {\bf n} \sum_{\hat e}
| \sum_i   
\frac{
(p_i\cdot {\hat e}) \epsilon_i
}
{
({\bf p}_i \cdot {\hat n} - p_{0i})}|^2
\end{equation}
where $p_i$ and ${\hat e}$ are  the 4-momenta and polarization of the  incoming and outgoing particles, $\epsilon_i=\pm 1$, for incoming particles or antiparticles; 
  $\epsilon_s $ an energy scale valid for single soft photon emission, to be determined to the order of precision in the perturbation treatment of the process to be studied.
The function $\beta$ 
was shown to be  a relativistic invariant  \cite{Etim:1967mxo} and   its  expression  in terms of the Mandelstam variables $s,t,u$ of 
two charged particle scattering  can be written as  \cite{Pancheri-Srivastava:1973wxx}:
 
\begin{equation}
\label{eq:beta}
\beta=\beta(s, t, u)=\frac{2e^2}{\pi }[-I_{12}+I_{13}
+I_{14} -2] 
\end{equation}
where 
\begin{equation}
I_{ij}=2 (p_i\cdot p_j) \int_0^1 \frac{dy}{2y(1-y) [(p_i\cdot p_j)-m_e^2]+m_e^2}
\end{equation}
with   the high energy limit  $s$, $u \rightarrow  \infty$, $-t \rightarrow 0$, $\beta \rightarrow -t/m_e$, which follows from the existence of the constant term in Eq.~(\ref{eq:beta}).

Following the steps taken from Eq.~[11] through Eq.~ [17] of the EPT paper 
 \cite{Etim:1967mxo},
  the analyticity properties of $h(t)$ in the lower half of the $t$-plane resulting from the constraint that $\omega \ge 0$, lead to
\begin{equation}
N(\beta) dP(\omega)= \beta \frac{d\omega}{\omega} (\frac{\omega}{E})^{\beta} \ \ \ \ \  for \  \omega< E
\label{eq:dPomega}
\end{equation}
with the normalization factor given by 
\begin{equation}
N(\beta)=\frac{\int_0^\infty dP(\omega)}{\int_0^E dP(\omega)} = \gamma^\beta \Gamma(1+\beta)
\end{equation}
which one obtains following the procedure outlined in Appendix III of  the  paper. 
\subsection{The zero momentum mode} 
\label{sec:Palumbo}

In this section we  return to Eq.~(\ref{eq:prob4discrete}) and  discuss  the  separation of  the zero momentum mode   from the continuum, 
in Abelian gauge theories
for  different  boundary conditions \cite{Palumbo:1983aa}.

Up to Eq.~(\ref{eq:prob4discrete}), the method developed to obtain the energy-momentum distribution $K_\mu$ is a  classical statistical mechanics exercise.   Going further requires
  an  expression for the average number of photons of momentum $\bf k$ and the choice of the boundary  conditions imposed upon the field. Before taking  the continuum limit, it has been suggested \cite{Palumbo:1983aa} to  separate the zero mode from the others. Let  the quantization volume be  $V=L^3$, and introduce $\mu$, a fictitious photon mass, in light of  eventually take the limit $L \rightarrow \infty $ and $\mu \rightarrow 0$.  Separating the zero momentum mode of energy $\omega_0$  from all the other modes, we write
\begin{eqnarray}
\label{eq:hdt}
h(t)=n_0 [1-e^{-i\omega_0 t}] + {\bar h}(t)= n_0[1-e^{-i\omega_0 t}]  +\nonumber\\ 
+ \beta \int_o^E \frac{dk}{k} [1-e^{-ikt}];\nonumber\\
\end{eqnarray}
with the photon mass now safely taken to be zero in the integral defining  ${\bar h}(t)$. For the zero mode, the $\mu \rightarrow 0$ limit  require more attention. One has
\begin{equation}
n_0 [1-e^{-i\omega_0 t}] \approx i n_0 \omega_0 t\rightarrow  i n_0\mu t  \equiv iW_0 t \label{eq:W0}
\end{equation}
We see that the zero momentum mode can introduce  a new energy scale $W_0$, namely
\begin{equation}
\label{eq:W0new}
W_0= \frac{4 \pi e^2 \mu }{L^3 \mu^3} \times \frac{1}{2}  |\sum_i\epsilon_i \vec{v}_i|^2\equiv G(\mu) F(E,m)
\end{equation}
{with the  finite dimensionless function  $F(E,m)=|\sum_i \epsilon_i {\bf v_i} |^2/2$ depending 
 on   mass and energy of the emitting particles.

The overall energy distribution can now be written 
\begin{equation}
dP(\omega)=\frac{d \omega}{2\pi} \int \  dt \ e^{i(\omega -W_0) t -{\bar h}(t)}
\end{equation}
The question  is how to take   the continuum limit, $L\rightarrow \infty$ and $\mu \rightarrow 0$ and, accordingly whether   the energy $W_0$ can    be finite. Various cases can be considered, in correspondence with different boundary conditions, as will be discussed in a separate publication. Here we only note that if the limit to  be taken is $\lim_{L\rightarrow \infty, \mu\rightarrow 0} L^3 \mu^3 = finite $, then $W_0=0$ in QED
but  $W_0\neq 0$  can happen if   $e^2 \mu\rightarrow finite$ in this limit.
In such  case the energy distribution would receive an extra contribution to the usual QED expression. Following the same derivation as in \cite{Etim:1967mxo} - based on  the analyticity properties of the energy distribution for extracting $dP(\omega)$ for $\omega\le E$, one obtains 
\begin{equation}
(\frac{\omega}{E})^\beta \rightarrow  (\frac{\omega}{E})^\beta \ (1- \frac{W_0}{\omega})^\beta 
\label{eq:PP}
\end{equation} 
 It should be noticed that the extra factor $(1- \frac{W_0}{\omega})^\beta $ is  regulated by the continuum contribution through the $\beta$ exponent.  

\section{How about QCD?}
The interest in the separation between the zero momentum from the continuum arises in QCD, in particular for the case of the transverse momentum distribution of the emitted radiation. Integrating the four momentum distribution over the energy and the longitudinal momentum variable, one can write the overall $d^2P({\bf K}_\perp)$ distribution  
  \cite{PancheriSrivastava:1976tm} as follows:
\begin{equation}
d^2P({\bf K}_\perp)=\frac{1}{(2\pi)^2}
 \int d^2{\bf b} 
 e^{
- i{\bf K}_\perp \cdot {\bf b} - \sum_{\bf k} \bar{n}_{\bf k} 
[ 
1 -e^{i {\bf k}_t \cdot {\bf b}}
]  } 
\end{equation}
where $\bar{n}_{\bf k}$ is the average number of soft gluons emitted with momentum $\bfk$.
Taking a straightforward  continuum limit, this expression leads to the  expression for soft gluon transverse momentum distribution 
\begin{equation}
 d^2P({\bf K}_\perp)=\frac{1}{(2\pi)^2}
 \int d^2{\bf b}  e^{- i{\bf K}_\perp \cdot {\bf b} -h(\bfb)}
\end{equation}
 with
\begin{equation}
\label{eq:hcontinuum}
h( b) =\int ^{qmax} d^2 \bfk_t  \frac{\alpha_s(k_t)}{k_t^2} \ln({2 qmax/k_t})  
[ 
1 -e^{i {\bf k}_t \cdot {\bf b}}
]  
\end{equation}
with $qmax$ an upper limit of integration which depends on the process under consideration. The still unknown  infrared behaviour of $\alpha_s$  
led  the lower limit of integration to be a scale $\Lambda \neq 0$ \cite{Dokshitzer:1978yd,Parisi:1979se,Curci:1979sk}, introducing  an intrinsic transverse momentum  for phenomenological applications{\cite{CMS:2024goo}. In our phenomenology of hadronic cross-sections, we have 
  set the  lower limit to be zero, introducing a singular but integrable behaviour for $\alpha_s$ in the infrared region as highlighted in the Introduction,  through Eq.~(\ref{eq:aphaBN}). As a result,    the integration is now dominated by a power law behaviour of $\alpha_s$ for $k_t\lesssim \Lambda$, and morphs into the asymptotic freedom expression beyond it.   More about this issue will be discussed 
   in Sect.~\ref{sec:alphas}. 

However, just as in the case of the energy distribution that  we have previously discussed, care is needed  in taking the continuum limit, and one must first separate out the zero momentum mode, with the result that the distribution now takes the form  

\begin{equation}
\label{eq:cutoff}
d^2P({\bf K}_\perp)=\frac{1}{(2\pi)^2}
 \int d^2{\bf b} 
 e^{
- i{\bf K}_\perp \cdot {\bf b} - \nu^2 b^2 - h_{continuum}({ b})}
\end{equation}
with
\begin{equation}
h_{continuum}(b)=\int_0^{qmax} d^3 \bar{n}_{\bf k} 
[1-e^{i {\bf k}_t \cdot {\bf b}}] 
\label{eq:hcont2}
\end{equation}
given in Eq.~\ref{eq:hcontinuum}}.
In Eq.~\ref{eq:cutoff} we argue that a cut off in impact space is developed through the   zero momentum mode ${\bf n}_0[1-e^{i {\bf k}_t \cdot {\bf b}}]\approx {\bf n}_0 [-i {\bf k}_t \cdot {\bf b}] + {\bf n}_0 \mu^2 b^2 +...\approx \nu^2 b^2$, 
 as the first term in the square bracket is killed by the integration over all directions, and only the second  term remains.
 The coefficient  $\nu^2=\mu^2 {\bf n}_0 $ would  be calculated by taking the zero mode limit of $\bar{n}_{\bf k}$ and is the analogue of the parameter $W_0$ discussed in the previous section for the energy distribution in  an Abelian theory.

We see that the Touschek resummation procedure, that began  with   statistical mechanics manipulations,  can be applied to go beyond the derivation of the well known exponentiation of infrared corrections. In this section, we have used  it to explore the possibility of the appearance of a cut-off in impact parameter space arising from the zero momentum mode in QCD. 
Whether such a term survives and manifests itself as the origin of the forward slope of the elastic differential $p-p/{\bar p}$ cross-ssection would depend upon whether the cut-off $\nu\ne 0$ in the zero momentum limit.  

 Thus the question is not only to  perform the integral for the continuum in  Eq.~(\ref{eq:hcont2}) in the unknown infrared region, but also to examine possible limits   of the cut-off scale $\nu$ in a theory such as QCD, 
inspecting the    zero momentum mode. These could be two different regimes, and need not require the same treatment. 
 

In previous publications we have in fact  proposed to evaluate  $\bar{n}_{\bf k}$ using a singular but integral expression for the strong coupling constant and we shall discuss it in Sect.~\ref{sec:alphas}, which  we now turn to.

\section{ Modelling  the  strong coupling constant in soft gluon resummation}
 \label{sec:alphas}
 \subsection{Models for $\alpha_s$ in the continuum}

 The ansatz in Eq.~(\ref{eq:aphaBN}), on which  we relied for our previous phenomenology for the hadronic cross-section,  is however unsatisfactory, as it introduced a  parameter $1/2 <p<1$. Such parameter  was  physically   justified   \cite{Grau:1999em}, as being  related to a confining potential $V(r) \approx r^{2p-1}$, but was otherwise an unknown number. 
To eliminate such  an extra parameter,  while, at the same time interpolating between the infrared and the asymptotically free regime,  as requested in our BN model \cite{Pancheri:2016yel}, we propose the following expression for the coupling :
 \begin{equation}
\alpha_s^{LO}(Q^2)=\frac{1}{
\ln[
1+
(
\frac{Q^2}{\Lambda^2})^{b_0}
]
} \ \ \ \ \ \ \ \ \ \ b_0=\frac{11 - 2 N_f/N_c}{4 \pi} \label{eq:alphasnew}
\end{equation}
namely we identify the unknown parameter $p$ with 
 $b_0$, thus simplifying  the expression of Eq.~(\ref{eq:aphaBN}),  with a $Q^2$ dependence  determined  through the anomalous dimensions.  A plot of the $Q^2$ behaviour of this function is shown in Fig.~\ref{fig:alphasfrom1403} \cite{Pancheri:2014rga}. The plot is obtained for the case of 3 flavours and $\Lambda=$100 MeV, a somewhat low value as compared with other current determinations. In our previous phenomenology, such a value was the one appropriate for a good description of  the total cross-section through QCD mini-jets and   infrared soft gluon corrections, using Eq.~(\ref{eq:aphaBN}). With such value of the $\Lambda$ parameter and $N_f=5$, we find $\alpha_s^{LO}(M_Z) =0.112$,  in good agreement with present determinations \cite{ParticleDataGroup:2024cfk}. The figure shows that  our proposal for a singular but integrable  $\alpha_s$ detaches from the asymptotic freedom curve  when   $\alpha_s \approx 0.7-0.8 $.
\begin{figure}[htb]
\centering
\includegraphics[width=8.5cm]{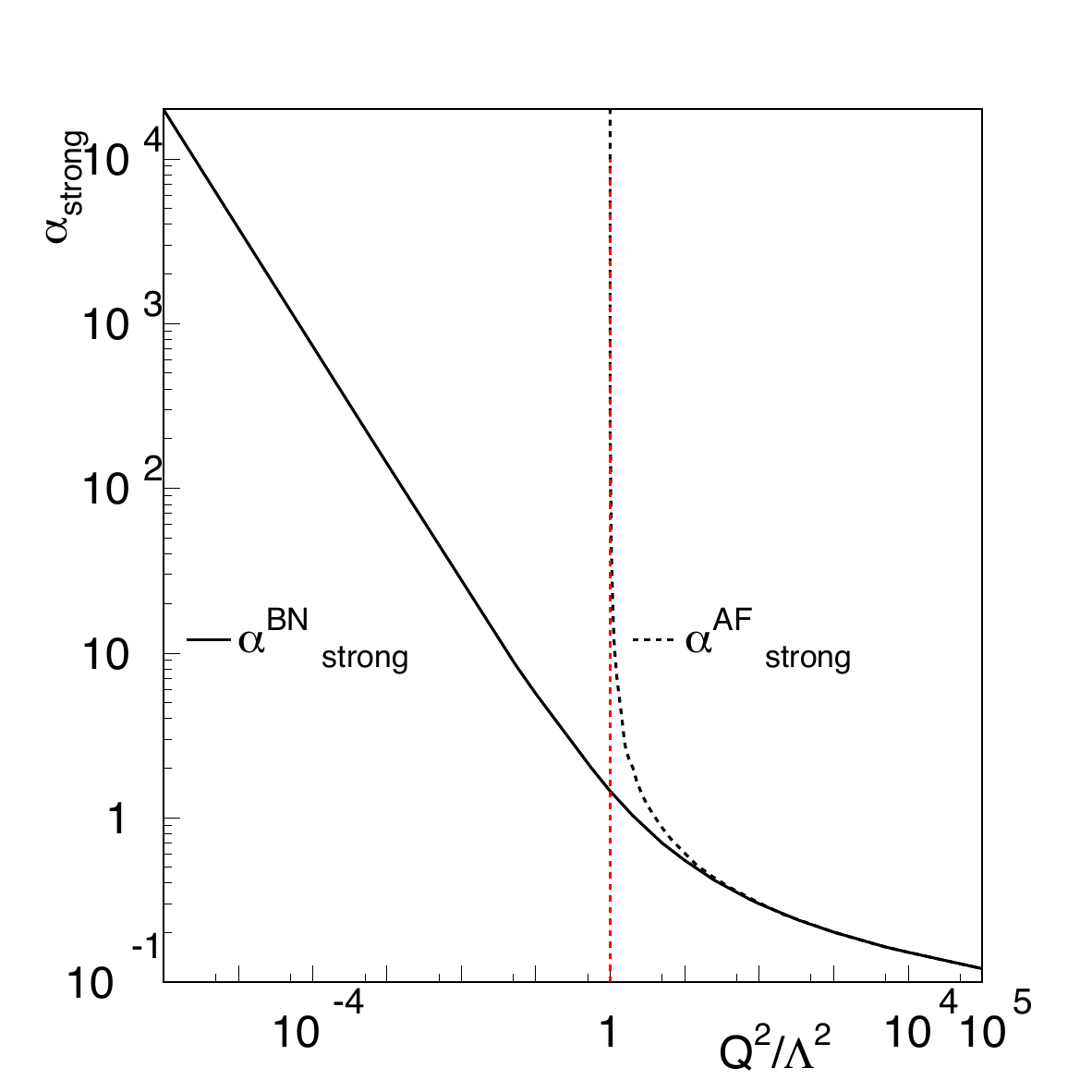}
\caption{The figure \cite{Pancheri:2014rga} shows the expression for $\alpha_s$ from Eq.~(\ref{eq:alphasnew})  and its comparison with the asymptotic freedom expression, for the case $N_f=3$ and $\Lambda=$100 MeV. }
\label{fig:alphasfrom1403}
 \end{figure}
 Comparison of the above expression with data from Jade, LEPII and LHC \cite{CMS:2014mna} is seen in Fig.~\ref{fig:alphaswithdata}.
\begin{figure}
 \centering
\includegraphics[width=8cm]{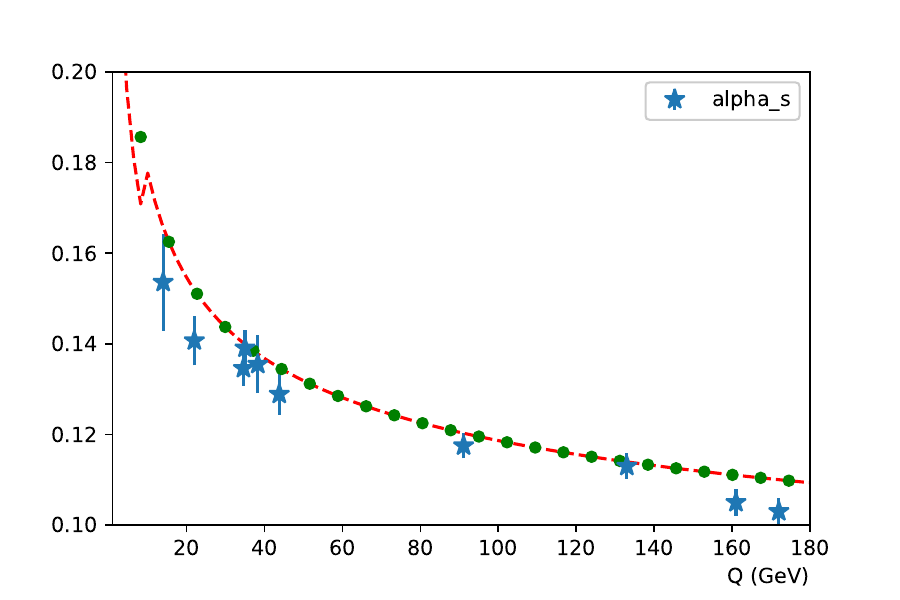}
\caption{The expression for $\alpha_ s$ as proposed in Eq.~(\ref{eq:alphasnew})   is compared with  experimental data  (green stars)  \cite{CMS:2014mna}. 
The red dashes correspond to $\Lambda = 100$  MeV and values of $N_f$
depending on the opening of the energy threshold, the green
dots are for $N_f = 5$. The gap corresponds to the threshold $N_f= 4 \rightarrow  5$.}
\label{fig:alphaswithdata}
 \end{figure}

 \subsection{Limiting behaviour of  $\alpha_s$}
 As discussed in Sec(\ref{sec:alphas}), the simplest \& most economical hypothesis for a confining $\alpha_s(Q^2)$
in consonance at the same time with asymptotic freedom (AF) [at leading order] is given by
\begin{equation}
\label{1}
\alpha_s(Q^2) = \frac{1}{ln [1 + (\frac{Q^2}{\Lambda^2})^{b_o}]}, 
\end{equation}
In an effort to find the renormalization group (RG) $\beta$- function that corresponds to Eq.(\ref{1}), we can first write 
\begin{equation}
\label{3}
\frac{1}{\alpha} = ln (1 + \eta); \eta = (\frac{Q^2}{\Lambda^2})^{b_o},
\end{equation}
and then write its derivative w.r.t. $\eta$ to read
\begin{eqnarray}
\label{4}
\frac{d\alpha}{d ln\eta} = \eta \frac{d\alpha}{d \eta} = - \alpha^2 \frac{\eta}{1 + \eta}\\
= - \alpha^2 [ 1 - \frac{1}{1 + \eta}] =  - \alpha^2 [ 1 - e^{-1/\alpha}]
\end{eqnarray}
having put $\alpha_s=\alpha$, for simplicity. Let us define $ln \eta = b_o t$. Then, with the standard definition of the RG beta function, 
\begin{equation}
\label{5}
\frac{d\alpha}{d t} = \beta (\alpha),
\end{equation}
corresponding to Eq.(\ref{1}) reads
\begin{equation}
\label{6}
\beta(\alpha) = -b_o \alpha^2 [1 - e^{-1/\alpha}]. 
\end{equation}
We note the following pleasing features of this beta function:  
\begin{eqnarray}
\label{7}
{\rm (i):}\ \beta(\alpha) < 0\\
{\rm (ii):}\ \beta \to -b_o \alpha^2\ {\rm for}\ \alpha < < 1; {\rm UV\ region}\\
{\rm (iii)}\ \beta \to -b_o \alpha\ {\rm for}\ \alpha > > 1; {\rm IR\ region}\\.
\end{eqnarray}
It may be of further interest to note that the dielectric function $\epsilon = n^2$,
where $n$ is the refractive index, defined as usual
\begin{equation}
\label{8}
\alpha = \frac{1}{\epsilon} = \frac{1}{n^2},
\end{equation}
the  ``beta-function'' corresponding to $n$, the refractive index in this model, has a scaling property.
\begin{eqnarray}
\label{9}
\frac{dn}{dt} = - \frac{\alpha^{-3/2}}{2} \frac{d\alpha}{d t} \equiv\ {\bar \beta}(n)\\ 
{\bar \beta}(n) = \frac{b_o}{2} [\frac{1 - e^{-n^2}}{n}]
\end{eqnarray}
Thus
\begin{eqnarray}
\label{10}
{\bar \beta}(n) \to (\frac{b_o}{2}) n \to 0\ {\rm for}\ n \to 0\ {\rm IR\ region};\\
{\bar \beta}(n) \to (\frac{b_o}{2}) \frac{1}{n} \to 0\ {\rm for}\ n \to \infty\ {\rm UV\ region}.
\end{eqnarray}
Hence, in this model, ${\bar \beta}(n)$ tends to zero both in the IR \& the UV regions, always remaining positive within the
finite domain. Moreover, it has an asymptotic $n \leftrightarrow 1/n$ symmetry as one goes from the IR to the UV region.  
\begin{equation}
\label{11}
{\bar \beta}(n) \leftrightarrow {\bar \beta}(1/n)\ {\rm as}\ n \to 0. 
\end{equation}
This interesting symmetry  appears to be a duality between strong and weak coupling similar to that conjectured from ADS/CFT  {\cite{Tan:2000sw,Brower:2008cy}}.

It should also be noted that as ${\bar \beta}(n)$ goes to zero both for $n \to 0$ as well as $n \to \infty$, it has a maximum.
Setting the derivative with respect to $n$ equal zero at $n = {\bar n}$ in Eq.(\ref{9}), we find the transcendental equation
for ${\bar n}$: 
\begin{equation}
\label{12}
e^{-{\bar n}^2} = \frac{1}{[1 + 2{\bar n}^2]}, 
\end{equation}
whose numerical solution is
\begin{equation}
\label{13}
[{\bar \beta}(n)]^2 \approx\ 1.275\ \ {\rm that\ corresponds\ to}\ {\bar \alpha} \approx\ 0.784,
\end{equation}
remarkably similar to Gribov's critical coupling for $SU(3)_c$, (see, Eq.(3) on page 344 of \cite{Gribov:2002}):
\begin{equation}
\label{14}
\alpha_c = \frac{2}{3 - (1/3)} = \frac{3}{4} 
\end{equation}

According to Gribov, the critical coupling above ($\alpha_s=3/4$ for $SU(3)_c$) makes the color Coulomb potential
between two colored quarks $U(r)= (4/3)(\hbar c) (\alpha_s/r)$ to take on the {\it magic value} $(\hbar c)/r$, exactly the value in QED when the nuclear charge $Z=137$ so that $Z\alpha_{QED}=1$ renders a  nucleus unstable to decay into an {\it ion} with nuclear charge $Z=136$ and a positron. In QCD, according to Gribov, when $\alpha_s$ becomes $3/4$ (or, higher) the ($q\bar{q}$) state manifests itself as a colorless meson.

\section{\bf Gribov dynamics with singular $\alpha_s(Q^2)$}
In the collection of papers [See \cite{Gribov:2002}: Collected Works, specially Chapter IV in the book, ``Gauge Theories \& Quark Confinement''], Gribov  writes down a set of equations for the quark Green's function. He assumes that there exists an IR region where $\alpha_s > \alpha_{critical}$.
Essentially, his work is focussed on an IR ``frozen'' $\alpha_s$ above the critical value. [See, his Eq.(92) in Chapter IV]. This is sufficient for him to obtain a phase transition, spontaneous breakdown of the chiral limit, the pion Goldstone modes etc. \cite{Gotsman:2020ryd} A tour de force indeed.\\
What we would like to do is to go beyond and ask the question: what happens for a non-frozen, IR singular but integrable $\alpha_s$ of the type discussed in the Introduction? This interesting region is not covered by Gribov. Let us condense the argument. Gribov defines the couplings ({\it Warning: the
$\beta$ below is not the usual $\beta$ such as in the previous section})
\begin{equation}
\label{2.1}
\beta = 1 - g;\ g = \frac{4 \alpha_s}{3 \pi}, 
\end{equation}
with $g > g_c$. Since his $g$ never gets close to 1, $\beta$ never hits zero. But look at his dynamical equations for the quark Green's function [below $A_\mu$ is not the vector potential but is dynamically generated]:
\begin{equation}
\label{2.2}
[\frac{\partial}{\partial q_\mu} - A^\mu(q)] G^{-1}(q) = 0;\ A^\mu(q) = [\frac{\partial}{\partial q_\mu} G^{-1}(q)] G(q) 
\end{equation}
and [with derivatives understood to be with respect to $q_\mu$]
\begin{equation}
\label{2.3}
\partial^\mu A_\mu(q) = - \beta(q) A_\mu(q) A^\mu(q). 
\end{equation}
For an IR singular $g$, $\beta$ can become zero (or even negative), cases {\bf not} considered by Gribov.  But when $\beta(q = q_o) = 0$, the rhs of Eq.(\ref{2.3}) is zero, it  appears that one has an interesting chiral invariant phase with $m(q_o)$ driven to zero. However, this value is unstable. All we can say for the moment is that it looks very attractive, exciting and needs further investigation. For example, the quark mass in the IR region seems to inherit power law growth -an anomalous dimension- given by $b_o$. \\

\section{\bf Dispersion relation \& a sum rule for the color refractive index \label{index} }
\noindent
To go beyond a specific model and discuss the general case, the time honored approach is to employ analyticity and write a dispersion relation for $\alpha_s(s)$ with a right-hand branch cut for $s\geq 0$; see:
\cite{Srivastava:2001ts,Milton:1996fc,Solovtsov:1999in,Shirkov:2006gv,Srivastava:2008pv,Malaspina:2024}. In these previous works, dispersion relations for $\alpha_s$ or for $\epsilon$ were employed, both requiring one subtraction. Here we employ a dispersion relation for the color refractive index 
$n(s)= \sqrt{\epsilon(s)}=1/\sqrt{\alpha_s(s)}$, that should have the same domain of analyticity as $\alpha_s$
provided in the space-like region ($s=-Q^2<0$), $\alpha_s(Q^2)$ does not vanish -a natural requirement for a coupling constant. Also, such a dispersion relation should require {\it no} subtraction under the hypothesis that (i) $\alpha_s(0)$ is either frozen (that is, it is a finite constant) or it diverges so that, $n(0)$ is finite or zero and      
(ii) for large $s$, asymptotic freedom prevails and thus $\Im m\ n(s) \to 0^-$ as $s\to \infty$.\\
Let us consider the refractive index $n(Q^2)$ in the space-like region $s=-Q^2<0$. Normalizing it at the
QCD scale $Q^2=\Lambda^2$, we have
\begin{eqnarray}
\label{i1}
n(Q^2)= n(\Lambda^2) + \frac{(\Lambda^2-Q^2)}{\pi} \int_o^\infty \frac{ds\ 
 \Im m\ n(s)}{(s+Q^2)(s+\Lambda^2)}; \nonumber\\  
 (\Im m\ n(s) < 0; s>0);\nonumber\\
{\rm Thus:}\ n(Q^2>\Lambda^2)> n(\Lambda^2)\nonumber\\  
\&\ \alpha_s(Q^2>\Lambda^2) < \alpha_s(\Lambda^2); {\rm (asymptotic\ freedom)};\nonumber\\
{\rm In\ the\ IR\ region:}\ n(Q^2<\Lambda^2) < n(\Lambda^2)\nonumber\\
\&\ \alpha_s(Q^2<\Lambda^2)>\alpha_s(\Lambda^2).
\end{eqnarray} 
Eq.(\ref{i1}) is satisfactory in that the fall-off of $\alpha_s(Q^2)$ for large $Q^2$ makes asymptotic freedom  evident and its rise in the IR region ($Q^2<<\Lambda^2$) bodes well for reaching or, even overreaching, the Gribov critical value $\alpha_s(Q^2<\Lambda^2)=3/4$, as discussed in the last section.\\
If indeed $n(0)=0$ (that is $\alpha_s(0)\to \infty$), then we have the sum rule
\begin{eqnarray}
\label{i2}
n(\Lambda^2) = \frac{\Lambda^2}{\pi} \int_o^\infty \frac{ds\ [-\Im m\ n(s)]}{s(s+\Lambda^2)}
\end{eqnarray}  
To delineate further between a finite versus a divergent $\alpha_s(0)$, we show that only a divergent $\alpha_s(0)$ is consistent with the asymptotic duality for $\bar{\beta}(Q^2)$, that was discussed earlier through a specific model - see Sec.(\ref{alphass}). The essential ingredient in obtaining the sought after asymptotic duality $\bar{\beta}_{IR}(Q^2)\leftrightarrow \bar{\beta}_{UV}$ as $n\leftrightarrow 1/n$ is that $n_{IR}(Q^2)$ vanish as a power-law  $(Q^2/\Lambda^2)^q$ as $Q^2 \to 0$; only then, it can be matched with its AF logarithmic behavior. While the argument is more general, for simplicity we shall show it here only for the lowest perturbative order:
\begin{eqnarray}
\label{i3}
n_{IR}(Q^2)\equiv n_o (\frac{Q^2}{\Lambda^2})^q;\nonumber\\
\bar{\beta}_{IR} = q\ n;(i);\nonumber\\
n_{UV}(Q^2)\equiv n_\infty [ln(\frac{Q^2}{\Lambda^2})]^{1/2};\nonumber\\  
\bar{\beta}_{UV}= (\frac{n_\infty^2}{2})\times
(\frac{1}{n}) = (\frac{b_o}{2}) \times \frac{1}{n};(ii);\nonumber\\
{\rm Thus,\ equality\ between\ (i)\ and\ (ii)\ implies}: q= \frac{b_o}{2};
\end{eqnarray}
exactly as we found in Eqs.(\ref{10} \&\ \ref{11}) for the specific model considered therein.\\

\section{\bf Conclusions \label{conc}} 
\noindent
 We have  approached the infrared region in QCD, both in the continuum and the zero momentum point. After recapitulating previous work in QED in a formalism developed by Bruno Touschek which had been  applied to investigate a  zero momentum mode in Abelian gauge theories, we have extended Touschek's approach to make the ansatz that could lead to a cut-off  in impact parameter space in parton-parton collisions and  shed light on the origin of the forward peak in hadronic collisions. 
    
Breaking with tradition,  we have considered dispersion relations for the color refractive index $n(s)= \sqrt{\epsilon(s)}=1/\sqrt{\alpha_s(s)}$ -for which no subtractions are needed- and a sum rule was derived for a divergent (but integrable $\alpha_s$ \cite{Grau:1999em}). It was also deduced that under the same hypothesis, an asymptotic duality ($n \leftrightarrow 1/n$) exists that in addition guarantees the integrability condition previously assumed in reference \cite{Grau:2009qx}. While our explicit expressions have been written down for 1-loop, the reader is encouraged to extend the formalism to higher loops. 

{We thank Fabrizio Palumbo, Simone Pacetti, L. Pierini  and A. Grau for their   contribution to discussions and interest in this problem.}

\bibliography{zeromode}

\end{document}